\documentclass[aps,pra,twocolumn,floatfix]{revtex4}
\usepackage{bm,dcolumn,amsmath,graphicx}
\begin{document}
\title{Study of highly-charged Ag-like and In-like ions for the development of atomic clocks and  search for $\alpha$-variation}
\author{M. S. Safronova$^{1,2}$}
\author{V. A. Dzuba$^{3}$}
\author{V. V. Flambaum$^{3}$}
\author{U. I. Safronova$^{4,5}$}
\author{S. G. Porsev$^{1,6}$}
\author{M. G. Kozlov$^{6,7}$}

\affiliation {$^1$University of Delaware, Newark, Delaware, USA}
\affiliation {$^2$Joint Quantum Institute, NIST and the University of Maryland, College Park, Maryland, USA}
\affiliation {$^3$The University of New South Wales, Sydney, Australia}
\affiliation {$^4$University of Nevada, Reno, Nevada, USA}
\affiliation {$^5$University of Notre Dame, Notre Dame, Indiana, USA}
\affiliation {$^6$Petersburg Nuclear Physics Institute, Gatchina,  Russia}
\affiliation {$^7$St.\ Petersburg Electrotechnical University ``LETI'', St.\ Petersburg, Russia}
\date{\today}

\begin{abstract}
We carried out detailed high-precision study of Ag-like Nd$^{13+}$, Sm$^{15+}$ and In-like Ce$^{9+}$, Pr$^{10+}$, Nd$^{11+}$, Sm$^{13+}$, Eu$^{14+}$ highly-charged ions. These ions were identified to be of particular interest to the development of ultra-accurate atomic clocks, search for variation of the fine-structure constant $\alpha$, and
quantum information [Safronova et. al. Phys. Rev. Lett. 113, 030801  (2014)]. Relativistic linearized coupled-cluster method was used
for Ag-like ion calculations, and a hybrid approach that combines configuration interaction and a variant of the coupled-cluster method was used for the In-like ion calculations. Breit and  QED corrections were included.  Energies, transition wavelengths, electric-dipole, electric-quadrupole, electric-octupole, magnetic-dipole, magnetic-quadrupole, magnetic-octupole reduced matrix elements, lifetimes, and sensitivity coefficients to $\alpha$-variation were calculated.
Detailed
study of various contributions was carried out to evaluate uncertainties of the final results.
Energies for several similar ``reference'' ions, where the experimental values are available were calculated and compared with experiment for further tests of the accuracy.
\end{abstract}
 \maketitle
\section{Introduction}
 The modern theories aimed at unifying gravitation with the three other fundamental interactions suggest
variation of the fundamental constants  in an expanding universe~\cite{Uza03}.
 A very large recent study of quasar absorption systems may indicate a spatial variation in the fine-structure constant $\alpha=e^2/\hbar c$ \cite{WebKinMur11}.
Spatial $\alpha$-variation hypothesis can be tested in terrestrial studies if sensitivity $\delta \alpha/\alpha \sim 10^{-19} \ {\rm yr}^{-1}$ is achieved~\cite{BerFla12,BerFlaOng13}.
Development of ultra-precise atomic clocks already allowed laboratory tests of the temporal $\alpha$-variation  at the present time. Different optical atomic clocks
use  transitions that have different  contributions of the relativistic corrections to frequencies. Therefore, comparison of these clocks can be used
to search for $\alpha$-variation.
 The most precise laboratory test of temporal $\alpha$-variation has been carried out at NIST \cite{RosHumSch08} by measuring the frequency
 ratio of Al$^+$ and Hg$^+$ optical atomic clocks
with a fractional uncertainty of $5.2\times10^{-17}$. Repeated measurements during the year yielded a constraint on the temporal variation of
$\alpha$  of $\dot{\alpha}/\alpha=(-1.6\pm 2.3)\times 10^{-17}$.
Development of ultra-precise atomic clocks is also essential for various other tests of fundamental physics, development of extremely sensitive quantum-based tools, very-long-baseline interferometry for telescope array synchronization, and tracking of deep-space probes \cite{HinShePhi13,BloNicWil14}.

Certain systems exhibit much higher sensitivity to the variation of $\alpha$
allowing more precise tests of the temporal variation and possible tests of the spatial variation of $\alpha$ \cite{WebKinMur11}.
 Selected transitions in highly-charged ions (HCI) were shown
to have  very large sensitivities to  $\alpha$-variation owing to  high nuclear charge $Z$, high ionization state, and
differences in the configuration composition of the corresponding states~\cite{BerDzuFla10,BerDzuFla11b}. While highly-charged ions  have very
large ionization energies, some of these systems have transitions that lie in the optical range due to level crossing.
 Moreover, these ions have very long-lived low-lying  metastable states which is a first requirement for the development of a frequency standard.
Highly-charged ions are less sensitive to external
perturbations than either neutral atoms or singly charged ions due
to more compact size of the electronic cloud.
As a result, some of the usual systematic clock uncertainties as well as decoherence processes in quantum information
applications may be suppressed.

Ag-like  Nd$^{13+}$ and Sm$^{15+}$ were proposed for the development of atomic clocks and
subsequent tests of the variation of the fine-structure constant in Refs.~\cite{DzuDerFla12,SafDzuFla14}.
 Detailed study of the potential clock uncertainties with these systems \cite{DerDzuFla12} have shown that
the fractional accuracy of the transition frequency in the clocks based on highly-charged ions can be smaller than
10$^{-19}$. Estimated sensitivity to the variation of $\alpha$ for highly-charged ions approaches
10$^{-20}$ per year \cite{DzuDerFla12}, which may allow for tests of spatial variation of the fine-structure constants that may be indicated by the observational studies \cite{WebKinMur11}. In-like Ce$^{9+}$, Pr$^{10+}$, Nd$^{11+}$ and Sm$^{13+}$
were proposed for the applications listed above in Ref.~\cite{SafDzuFla14}. Experimental work in HCIs requires knowledge
of many atomic properties of these systems, especially, wavelengths, transition rates, and lifetimes. To the best of our knowledge, no transition rates or lifetimes have been measured for any of the HCIs studied in this work. The energy levels
have only been measured for  Nd$^{13+}$, Sm$^{15+}$, and Ce$^{9+}$. No experimental data at all exist for Pr$^{10+}$, Nd$^{11+}$, Sm$^{13+}$, and Eu$^{14+}$.
Accurate  theoretical predictions of the transition wavelengths for these systems are particulary difficult owing to severe cancellations of upper and lower state energies, and we use the most high-precision methods available to perform the calculations.

In this work, we carried out detailed high-precision study of Ag-like Nd$^{13+}$ and Sm$^{15+}$ and In-like Ce$^{9+}$, Pr$^{10+}$, Nd$^{11+}$, Sm$^{13+}$, and Eu$^{14+}$ highly-charged ions using a relativistic linearized coupled-cluster method and a hybrid approach that combines configuration interaction and a variant of the coupled-cluster method. Breit and  QED corrections were included into the calculations.  Key results were presented in Ref.~\cite{SafDzuFla14}. Our calculations include energies, transition wavelengths, electric-dipole, electric-quadrupole, electric-octupole, magnetic-dipole, magnetic-quadrupole, magnetic-octupole transition rates, lifetimes, and sensitivity coefficients to $\alpha$-variation $q$  and $K$. We carried out extensive study of the uncertainties of our results. Three independent methods were used for the uncertainty studies:
\begin{itemize}
\item
Energies of Nd$^{13+}$, Sm$^{15+}$, and In-like Ce$^{9+}$ as well as several
similar ``reference'' ions Cs$^{6+}$, Ba$^{7+}$, and Ba$^{9+}$, where the
experimental values are available, were calculated and compared with experiment.
 \item For three of the In-like Ce$^{9+}$, Pr$^{10+}$, and Nd$^{11+}$ ``monovalent'' ions, both of the approaches used in this work are applicable so we were able to compare the various properties calculated with both methods to study the accuracy of the calculations.
  \item   Detailed
study of higher-order, Breit, QED, and higher partial wave contributions was carried out to evaluate uncertainties of the final results for each ion.
\end{itemize}
We start with the brief description of the methods used in this work in Section~\ref{method}.   The results for
Ag-like and In-like ions are presented in Sections ~\ref{ag} and \ref{in}, respectively.

\section{Methods}
\label{method}
We use two different relativistic high-precision approaches for all of the  calculations carried out in this work.
The first approach is the relativistic linearized coupled-cluster  method that includes all single, double, and partial
triple excitations (SDpT) of Dirac-Fock wave function~\cite{SafJoh08}. It is applicable only to monovalent systems, so we use it for the calculation of properties of Ag-like ions and those In-like ions that can be treated as monovalent systems. SDpT has been extremely
successfully in predicting properties of alkali-metal atoms and other monovalent ions \cite{SafJoh08}.
\subsection{Monovalent systems: all-order SDpT method}
The Ag-like ions have  a single valence electron
above the closed $1s^2 2s^2 2p^6 3s^2 3p^6 3d^{10} 4s^2 4p^6 4d^{10}$ core. It allows us to use  relativistic linearized coupled-cluster  method that includes all single, double, and partial
triple excitations of Dirac-Fock wave function. We refer the reader to the review \cite{SafJoh08} for detail description of the
method and its applications and give only a brief introduction to this approach.

The point of departure in all our calculations is the relativistic no-pair Hamiltonian $H=H_0+V_I$
 \cite{BroRav51} expressed for the case of \textit{frozen-core} $V^{N-1}$ Dirac-Fock potential as
\begin{align}
H_0 & = \sum_i \epsilon_i :a_i^\dagger a_i:\, , \label{np1}\\
V_I & = \frac{1}{2} \sum_{ijkl} g_{ijkl}  : a^\dagger_i a^\dagger_j a_l a_k :\, , \label{np2}
\end{align}
where $g_{ijkl}$ are two-particle matrix elements of the Coulomb
interaction, $\epsilon_i$ in Eq.~(\ref{np1}) is the eigenvalue
of the Dirac equation, $a^\dagger_i$, $a_i$ are creation and annihilation operators, and $:~:$
designate normal ordering of operators with respect to the core.

In the linearized coupled-cluster SDpT approach the atomic wave function of a
monovalent atom in a state $v$ is given by an expansion
\begin{eqnarray}
\lefteqn{ |\Psi_v \rangle = \left[ 1 + \sum_{ma} \, \rho_{ma}
a^\dagger_m a_a + \frac{1}{2} \sum_{mnab} \rho_{mnab} a^\dagger_m
a^\dagger_n a_b a_a
 \right. } \hspace{0.5in} \label{eq1} \nonumber \\
&& + \sum_{m \neq v} \rho_{mv} a^\dagger_m a_v + \sum_{mna}
\rho_{mnva} a^\dagger_m a^\dagger_n a_a a_v
\hspace{0.5in} \nonumber \\
&& + \left.
\frac{1}{6}\sum_{mnrab}
\rho_{mnrvab} a^\dagger_m a^\dagger_n a^\dagger_r a_b a_a a_v
\right]a^\dagger_v | \Psi_{C}\rangle
\,.
\end{eqnarray}
Here indices $a$ and $b$ range over all occupied core
states while the indices $m$, $n$, and $r$ range over all possible virtual
states,  and $| \Psi_{C}\rangle$ is the lowest-order frozen-core wave
function.  The quantities $\rho_{ma}$, $\rho_{mv}$ are
single-excitation coefficients for core and valence electrons;
$\rho_{mnab}$ and $\rho_{mnva}$ are core and valence
double-excitation coefficients, respectively. The triple excitations  $\rho_{mnrvab}$ are included perturbatively into the energy and single-valence excitation coefficient equations. In the single-double (SD)
variant  of the all-order method, only single and double
excitations are included.

The equations for the excitation coefficients $\rho$ and the correlation energy are derived by substituting the state vector
$|\Psi_v\rangle$ into the many-body Schr\"{o}dinger equation $H | \Psi_v\rangle=E| \Psi_v\rangle$. The resulting system of equations is solved iteratively until the correlation energy converges to required numerical accuracy.
This approach
includes dominant many-body perturbation theory (MBPT) terms to all orders because  every iteration
picks up correlation terms that correspond to the next order of
perturbation theory.

The matrix elements of any one-body operator $Z = \sum_{ij} z_{ij}\
a^\dagger_i a_j$, such as transition operators $Ek$ and $Mk$, $k = 1, 2, 3$ needed for this work,  are obtained within the framework of the all-order
method as
\begin{eqnarray}
Z_{wv} &=& \frac{\langle \Psi_w |Z| \Psi_v \rangle}{\sqrt{\langle \Psi_v
| \Psi_v \rangle \langle \Psi_w | \Psi_w \rangle}} \nonumber \\
       &=& \frac{z_{vw}+Z^{(a)}+\cdots+Z^{(t)}}{\sqrt{\langle \Psi_v
| \Psi_v \rangle \langle \Psi_w | \Psi_w \rangle}} ,  \label{eqr}
\end{eqnarray}
where $|\Psi_v\rangle$ and $|\Psi_w\rangle$ are given by the
expansion (\ref {eq1}) restricted to  SD approximation.  The terms $Z^{(a)}\cdots Z^{(t)}$ are linear
or quadratic functions of the excitation coefficients and $z_{wv}$ is the DF matrix element~\cite{BluJohLiu89}.
\begin{table}
\caption{\label{tab1} Second-order Coulomb correlation energy calculated with $l_{\rm max}=5$,
$l_{\rm max}=6$, and final extrapolated values (Final). The contributions of $l=6$ and $l>6$ are compared
in the last two columns. All values are in cm$^{-1}$.}
\begin{ruledtabular}
\begin{tabular}{lrrrrr}
\multicolumn{1}{c}{Level}&
\multicolumn{1}{c}{$l_{\rm max}=5$}&
\multicolumn{1}{c}{$l_{\rm max}=6$}&
\multicolumn{1}{c}{Final}&
\multicolumn{1}{c}{$l=6$}&
\multicolumn{1}{c}{$l>6$}\\
\hline
$   5s  $          &   -22140  &   -22402  &   -22672  &   -262    &   -270    \\
$   5p_{1/2}   $   &   -20451  &   -20710  &   -20989  &   -259    &   -278    \\
$   5p_{3/2}   $   &   -19136  &   -19375  &   -19632  &   -239    &   -257    \\
$   4f_{5/2}   $   &   -26474  &   -27931  &   -29418  &   -1456   &   -1488   \\
$   4f_{7/2}   $   &   -26044  &   -27493  &   -28971  &   -1449   &   -1478   \\
 \end{tabular}
\end{ruledtabular}
\end{table}

We use a
complete set of DF wave functions on a nonlinear grid generated using
B-splines constrained to a spherical cavity $R=60$~a.u. The basis set
consists of 50 splines of order 9 for each value of the relativistic
angular quantum number $\kappa$.

The Breit interaction is included in the construction of the basis set.  The QED radiative corrections to energy levels are included using the method described in \cite{FlaGin05}. The contribution of the QED corrections for the ions calculated in this work is only significant for the configurations that contain valence  $5s$ state. Therefore, the QED can be omitted for ions where none of the low-lying configurations  contain $5s$ valence state.

 The partial waves with $l_{max}=6$ are included in all internal summations over all excited states.
 We find that
inclusion of the higher partial waves with $l>6$ is very important for accurate description of the $4f$ states.
We use second-order perturbation theory  where we can extrapolate the result to $l_{max}=\infty$ to evaluate the contribution
of $l>6$. The results are
illustrated in  Table~\ref{tab1} where we list second-order Coulomb correlation energies calculated with $l_{\rm max}=5$ and $l_{\rm max}=6$, their difference which is the contribution of $l=6$, and the final results calculated with $l=10$ partial waves and extrapolated to account for $l>10$ contributions.  The difference of the final and  $l_{\rm max}=6$ results gives the contribution of the $l>6$ partial waves (last column). We find that it is remarkably close to the contribution of the $l=6$ partial wave. The second order dominates correlation energy, therefore, this empirical rule is expected to hold for the all-order correlation corrections. As a result, we estimate the effect of higher partial waves in all of the calculations in this work by carrying out the entire all-order calculation with $l_{\rm max}=5$ and $l_{\rm max}=6$
and adding the difference to the final result. We label this contribution ``Extrap'' in all tables below.
\begin{table*}
\caption{\label{tab-ag-like} Energies of Ag-like Ba$^{9+}$, Nd$^{13+}$, and
Sm$^{15+}$ ions relative to the ground state evaluated in the SDpT all-order
approximation (in cm$^{-1}$). Contributions from higher-order Coulomb correlation (above second-order MBPT),
estimated contributions of  higher partial waves (above $l>6$),
Breit interaction, and QED are given separately in columns HO, Extrap, Breit, and QED, respectively.
 Experimental results are from \cite{nist-web} for Ba$^{9+}$ and \cite{ag-like-81} for Nd$^{13+}$ and
Sm$^{15+}$. Differences with experiment are given in cm$^{-1}$ and \% in columns ``Diff.'' and ``Diff.\%''. Wavelengths for
transitions to the ground state are given in the last two columns in nm.}
\begin{ruledtabular}
\begin{tabular}{rrrrrrrrrrrrrr}
\multicolumn{1}{c}{Ion}&
\multicolumn{1}{c}{Level}&
 \multicolumn{1}{c}{Expt.}&  \multicolumn{1}{c}{Ref.~\cite{DzuDerFla12}}&    \multicolumn{1}{c}{MBPT2}&
\multicolumn{1}{c}{HO}&
\multicolumn{1}{c}{Extrap}&
\multicolumn{1}{c}{Breit}&
\multicolumn{1}{c}{QED}&
\multicolumn{1}{c}{Final}&
\multicolumn{1}{c}{Diff.}&
\multicolumn{1}{c}{Diff.\%}& \multicolumn{1}{c}{$\lambda_\textrm{th}$}&
 \multicolumn{1}{c}{$\lambda_\textrm{expt}$}\\
\hline   \\[-0.4pc]
 Ba$^{9+}$ &$5s_{1/2}$ &       0&     &     0&     0   &      0&      0& 0    &      0   &     &   &     & \\
           &$5p_{1/2}$ &  139348&     &140221&  -719   & -18   & 304   &-530  & 139258   &  90  &0.06\%&71.81&71.76\\
           &$5p_{3/2}$ &  166361&     &167744&  -950   & -2    & -24   &-483  & 166285   &  76  &0.05\%&60.14&60.11\\
           &$4f_{5/2}$ &  222558&     &224696&   139   &-1006  & -912  &-569  & 222350   &  208 &0.09\%&44.97&44.93\\
           &$4f_{7/2}$ &  224074&     &226433&   57    &-997   & -1079 &-560  & 223856   &  218 &0.10\%&44.67&44.63\\ [0.5pc]
Nd$^{13+}$ & $5s_{1/2}$&       0&    0&     0&   0   &     0 &      0&   0     &        0&    &        &      &         \\
           & $4f_{5/2}$&   55870&58897& 58596& 761   & -1247 &  -1421&   -983  &   55706&  164& 0.29\% & 179.5  &179.0  \\
           & $4f_{7/2}$&   60300&63613& 63429& 671   & -1238 &  -1767&  -961   &   60134&  166& 0.28\% & 166.3  &165.8  \\
           & $5p_{1/2}$&  185066&     &185876&-492   &   -33 &    560&   -883  &   185028&   38& 0.02\% & 54.05 & 54.03 \\
           & $5p_{3/2}$&  234864&     &236463&-74  7 &   -14 &    -14&   -801  &   234887&  -23&-0.01\% & 42.57 & 42.58  \\ [0.5pc]
Sm$^{15+}$ &$4f_{5/2}$ &       0&    0&     0   &    0 &    0&      0&     0&         0    &        &        &       &        \\
           &$4f_{7/2}$ &    6555& 6806&  6949   &  -92 &   9 & -454  &  31  &      6444 &      111&   1.69\%  &  1552     & 1526  \\
           &$5s_{1/2}$ &   60384&55675&  57100  &  -820&1316 & 1686  &1236  &     60517&     -133&  -0.22\%  & 165.2      & 165.6  \\
           &$5p_{1/2}$ &  268488&     &  266011 & -1218&1277 & 2398  & 138  &         268604&     -116&  -0.04\%  & 37.23 & 37.25  \\
           &$5p_{3/2}$ &  333203&     &  331659 & -1482&1297 & 1870  & 243  &         333385&     -182&  -0.05\%  &  29.99& 30.01   \\
\end{tabular}
\end{ruledtabular}
\end{table*}
\subsection{Multivalent systems: CI+all-order method}
The linearized coupled-cluster method used for Ag-like ions is not directly
applicable for systems with two or more valence electrons. We use a hybrid
method that combines the modified linearized single-double (SD)
coupled-cluster method with configuration approach developed in
\cite{Koz04,SafKozJoh09}. The CI many-electron wave function is obtained as a linear
combination of all distinct states of a given angular momentum $J$ and parity
\cite{DzuFlaKoz96b}:
\begin{equation}
\Psi_{J} = \sum_{i} c_{i} \Phi_i\,.
\end{equation}
Then, energies and wave functions of low-lying states are determined by
diagonalizing the effective Hamiltonian:
\begin{equation}
H^{\text{eff}}=H_{1} + H_2, \label{ham}
\end{equation}
where $H_{1}$ and $H_2$ represents the one-body and two-body parts of the Hamiltonian, respectively.  The
matrix elements and other properties, such as electric-multipole and magnetic-multipole transition matrix elements, can be determined using the resulting wave functions.

The CI + many-body perturbation theory (MBPT) approach developed in ~\cite{DzuFlaKoz96b}
allows one to incorporate core excitations in
the CI method by including  perturbation theory terms into an
effective Hamiltonian~(\ref{ham}). Then, the one-body part $H_1$ is modified to include
the correlation potential $\Sigma_1$ that accounts for part of the core-valence correlations:
\begin{equation}\label{H1eff}
H_1 \rightarrow H_1+\Sigma_1.
\end{equation} and
the two-body Coulomb interaction term $H_{2}$ is modified by including
the two-body part of core-valence interaction that represents
screening of the Coulomb interaction by valence electrons;
\begin{equation}\label{H2eff}
H_2 \rightarrow H_2+\Sigma_2.
\end{equation}
The CI method is then applied as
usual with the modified $H^\text{eff}$ to obtain improved energies
and wave functions.
 In the CI + all-order approach,
  the corrections to the effective Hamiltonian
  $\Sigma_1$ and $\Sigma_2$ are calculated using a modified version of the
  linearized coupled-cluster all-order method described above which
  allows to include dominant core and core-valence correlation
  corrections to the effective Hamiltonian to all orders. The detailed
  description of the CI+all-order method and all formulas are given in
  ~\cite{SafKozJoh09}.
  Since the
CI space includes only three valence electrons for In-like ions, it can be made essentially complete.
 The CI+all-order method yielded accurate wave functions for the calculations of
such atomic properties as lifetimes, polarizabilities, hyperfine structure
constants, etc. for a number of divalent systems and Tl
\cite{SafKozJoh09,SafKozCla11,PorSafKoz12,SafPorKoz12,SafKozCla12,SafKozSaf12,SafPorCla12}.
We refer the reader to Refs.~\cite{SafJoh08,SafJohSaf06,SafSaf11} and
\cite{SafKozJoh09,SafPorCla12,SafKozCla11,SafKozSaf12,SafPorKoz12,PorSafKoz12} for detailed descriptions of the
linearized coupled-cluster and CI+all-order methods, respectively.
We use both methods for In-like ions with the exception of Sm$^{13+}$ and Eu$^{14+}$ that have low-lying trivalent
 configurations, such as $4f^25s$ as an additional test of accuracy.

 As in the monovalent all-order method, we included the Breit interaction on the same
footing as the Coulomb interaction in the basis set, which incorporates higher-order Breit effects. The  Gaunt part of the Breit interaction is included in the CI.  The contribution of the $l>6$ partial waves is calculated as described above, i.e.
using the empirical result that  total $l>6$ extrapolated contribution is approximately equal to the $l=6$ contribution.
  To evaluate the uncertainty of our calculations, we carried out several calculations for each ion to separate  the contributions of the higher-order Coulomb correlation, Breit, QED, and $l>6$ higher partial waves. Several methods are developed to evaluate the uncertainties.

The sensitivity of the atomic transition frequency $\omega$
 to the variation of the fine-structure constant $\alpha$ can be quantified using a coefficient $q$ defined as
 \begin{equation}
 \omega(x)=\omega_0+qx,
 \end{equation}
  where
   \begin{equation}
 x=\left(\frac{\alpha}{\alpha_0}\right)^2-1.
 \end{equation}
In the equation above, the frequency $\omega_0$ corresponds to the value of the fine-structure constant $\alpha_0$ at some initial point in time. It is preferable to select transitions with significantly different values
of  $q$, since the ratio of
two frequencies, which is a dimensionless quantity, is studied over time in the experiment. Extra enhancement will be present if $q$ for these transitions have different signs. We also define a dimensionless  enhancement factor $K=2q/\omega$.

The   calculation of sensitivity coefficient $q$ requires a performance of three calculations with different values of $\alpha$ for every ion considered in this work.
 First, the calculation is carried out with the current $\textrm{CODATA}$ value of $\alpha$ ~\cite{MohTayNew11}. Next, two
 other calculations are performed with
  $\alpha^2$ varied by $\pm$1\%.   The value of $q$ is then determined as a numerical derivative
 \begin{equation}
 q=\frac{\omega(0.01)-\omega(-0.01)}{0.02},
 \end{equation}
 where $\omega(\pm0.01)$ are results of the calculations with $\alpha^2$ varied by $\pm$1\%, respectively. The other calculation (with CODATA value of $\alpha$)
 is used to verify that the change in $\omega$ is very close to linear.  We also carried out test calculation for one of the ions, Pr$^{10+}$, with changing $\alpha^2$ by $\pm5\%$
 and obtained results identical to the ones obtained with $\pm1\%$ change.

\begin{table}
\caption{\label{tab-ag-like-q} Energies and  sensitivity coefficients $q$
for Ag-like ions relative to the ground state evaluated in the SDpT all-order
approximation in cm$^{-1}$; $K=2q/\omega$ is the enhancement
factor. Lowest-order DF sensitivity coefficients $q$ are given for comparison.}
\begin{ruledtabular}
\begin{tabular}{rrrrrr}
\multicolumn{1}{c}{Ion}&
\multicolumn{1}{c}{Level}&
\multicolumn{1}{c}{Energy}&
\multicolumn{1}{c}{$q$ (DF)}&
\multicolumn{1}{c}{$q$ (SDpT)}&
\multicolumn{1}{c}{$K$}\\
\hline   \\[-0.4pc]
Nd$^{13+}$  &   $5s_{1/2}$  &   0       &       0   &   0       &\\
            &   $4f_{5/2}$  &   55706   &   102609  &   104229  &   3.7 \\
            &   $4f_{7/2}$  &   60134   &   106276  &   108243  &   3.6 \\
            &   $5p_{1/2}$  &   185028  &   16047   &   15953   &   0.2 \\
            &   $5p_{3/2}$  &   234887  &   71013   &   72079   &   0.6 \\ [0.5pc]
Sm$^{15+}$  &   $4f_{5/2}$  &   0       &       0   &   0       &       \\
            &   $4f_{7/2}$  &   6444    &   5536    &   5910    &   1.8 \\
            &   $5s_{1/2}$  &   60517   &   -132449 &   -134148 &   -4.4    \\
            &   $5p_{1/2}$  &   268604  &   -113153 &   -114999 &   -0.9    \\
            &   $5p_{3/2}$  &   333385  &   -40883  &   -41477  &   -0.2
\end{tabular}
\end{ruledtabular}
\end{table}

The lifetime of a state $a$ is calculated as
$$
\tau_a=\frac{1}{\sum_{b} A_{ab}}.
$$
The multipole transition rates $A_{ab}$ are determined using the formulas:
\begin{eqnarray}
 A(E1) &=& \frac{2.02613\times 10^{18}} {(2J_a+1)\lambda ^{3}} \, \, S(E1),\\
  A(M1) &=& \frac {2.69735\times 10^{13}}{(2J_a+1)\lambda ^{3}} \, \, S(M1),\\
   A(E2)& =& \frac{1.11995\times 10^{18}}{(2J_a+1)\lambda ^{5}} \, \, S(E2),\\
  A(M2)& =& \frac{1.49097 \times 10^{13}}{(2J_a+1)\lambda ^{5}} \, \, S(M2),\\
   A(E3)& = &\frac{3.14441 \times 10^{17}}{(2J_a+1)\lambda ^{7}} \, \, S(E3),\\
  A(M3) &= &\frac{4.18610 \times 10^{12}}{(2J_a+1)\lambda ^{7}} \, \, S(M3),
 \end{eqnarray}
where the wavelength $\lambda$ is in \AA~and the line strength $S$ is in atomic units.

\section{Ag-like ions}

\label{ag}
The $5s-4f$ level crossing (i.e. change of the level order) in Ag-like isoelectronic sequence happens from Nd$^{13+}$ to Sm$^{15+}$. The order of the first few levels for previous ions, such as Ba$^{9+}$ is $5s$, $5p$, and $4f$.  The ordering becomes $5s$, $4f$, $5p$ for  Nd$^{13+}$ and then finally switches to $4f$, $5s$, $5p$ for Sm$^{15+}$. The Pm$^{14+}$ has no stable isotopes and we do not list its energies here. However, we find that the $5s$ and $4f_{5/2}$ states are separated by only about 300~cm$^{-1}$.

We list the energies of Ag-like Ba$^{9+}$, Nd$^{13+}$, and Sm$^{15+}$ ions relative to the ground state evaluated in the SDpT all-order approximation  in Table~\ref{tab-ag-like} (in cm$^{-1}$). Since the experimental energies are available for Nd$^{13+}$ and
Sm$^{15+}$, Ag-like ions represent excellent benchmark systems for our calculations.
While Ba$^{9+}$ is not of practical interest for applications of this work,  experimental data are available
for Ag-like and In-like Ba ions. Therefore, we carried out calculations for these Ba ions to provide similar reference systems. The total of lowest-order DF and second order values are given in column labelled ``MBPT2''.
Contributions from higher-order Coulomb correlation,
estimated contributions of  higher partial waves (above $l>6$),
Breit interaction, and QED are given separately in columns HO, Extrap, Breit, and QED.
The higher-order corrections are calculated as the difference of the all-order and the second-order results. All all-order calculations include partial triple excitations as described above. The Breit contribution is calculated as the difference of the energies obtained with and without the inclusion of the Breit interactions. The QED radiative corrections to energy levels are included using the method described in \cite{FlaGin05}.

\begin{table*}
\caption{\label{tab-mult1} Lowest order ($Z^{\rm DF}$) and all-order ($Z^{\rm SDpT}$)  multipole matrix elements
$Ek$ and $Mk$, $k=1,2,3$ in a.u., transition rates $A$ (in s$^{-1}$), and lifetimes in Ag-like ions. Experimental energies from Ref.~\cite{ag-like-81} are used in calculation of transition rates. Energies (in cm$^{-1}$) and wavelengths (in nm) are listed for reference.
   The numbers in brackets represent powers of 10.}
\begin{ruledtabular}
\begin{tabular}{lrrrrrrrr}
\multicolumn{1}{c}{Level}& \multicolumn{2}{c}{Transition}&
\multicolumn{1}{c}{Energy}& \multicolumn{1}{c}{$\lambda$}&
\multicolumn{1}{c}{$Z^{\rm DF}$}& \multicolumn{1}{c}{$Z^{\rm
SDpT}$}&
\multicolumn{1}{c}{$A$}& \multicolumn{1}{c}{Lifetime} \\ \hline
 \multicolumn{9}{c}{  Ag-like Nd$^{13+}$}\\ \\[-0.6pc] $4f_{5/2}$ &$4f_{5/2}-5s_{1/2}$  & E3&   55870&   179.1 &   0.955&    0.922 & 7.568[-07]&\textbf{15.3~days}\\
           &$4f_{5/2}-5s_{1/2}$  & M2&   55870&   179.0 &   0.00004&    0.00038 & 1.987[-11]& \\[0.3pc]
 $4f_{7/2} $ &$4f_{7/2}-4f_{5/2}$&  M1 &   4430&    2257&     1.850&     1.850&     1.004&0.996~s  \\
 &$4f_{7/2}-4f_{5/2}$&           E2 &   4430&    2257&     0.320&     0.285&     1.936[-06]&\\
 &$4f_{7/2}-5s_{1/2}$&           E3 &   60300&   165.8 &   1.113&     1.076&     1.319[-06]&\\  [0.3pc]
  $5p_{1/2}$&$5p_{1/2}-5s_{1/2}$& E1& 185066&     54.03&     1.018&     0.873&     4.899[09]   &   0.204~ns\\  [0.3pc]
 $5p_{3/2}$ &$5p_{3/2}-5s_{1/2}$& E1& 234864&     42.58&     1.446&     1.245&     1.016[10]&  0.0984~ns\\  [0.3pc]
 \multicolumn{9}{c}{  Ag-like Sm$^{15+}$}\\ \\[-0.6pc]
 $4f_{7/2}$    &$4f_{7/2}-4f_{5/2}$ &  M1 &  6555&   1525.6&     1.850&     1.850&     3.251 &0.308~s\\
   &$4f_{7/2}-4f_{5/2}$ &  E2 &  6555&   1525.6&     0.256&     0.228&     8.801[-06]&\\       [0.3pc]
    $5s_{1/2}$  &$ 5s_{1/2}-4f_{5/2}$&  E3 &  60384&   165.6 &   0.676&    0.657 &  1.986[-06]&\textbf{ 3.62~days}\\
   &$ 5s_{1/2}-4f_{7/2}$&  E3 &  53829&   185.8 &   0.789&    0.768 &  1.214[-06]&\\
   &$ 5s_{1/2}-4f_{5/2}$&  M2 &  60384&   165.6 &   0.00004&   0.00025 &  3.643[-11]&\\       [0.3pc]
$5p_{1/2}$   &$5p_{1/2}-5s_{1/2}$ &  E1 &208104&     48.05&     0.940&     0.809&   5.978[09]&0.167~ns\\
$5p_{3/2}$   &$5p_{3/2}-5s_{1/2}$ &  E1 &272819&     36.65&     1.337&     1.153&   1.368[10]&0.0731~ns
   \end{tabular}
\end{ruledtabular}
\end{table*}

 Experimental results are from \cite{nist-web} for Ba$^{9+}$ and \cite{ag-like-81} for Nd$^{13+}$ and
Sm$^{15+}$. Difference with experiment is given in cm$^{-1}$ and \% in columns ``Diff.''  and ``Diff.\%''. Wavelengths for
transitions to the ground state are given in last two columns in nm. Our results are in excellent agrement with experimental data. We also include comparisons with recent correlation potential method results of Ref.~\cite{DzuDerFla12}. The table illustrates that inclusion of Breit, higher-order partial waves, and QED contributions is essential for achieving accurate  results.

 The sensitivity coefficients $q$ for Ag-like ions obtained as described in Section~\ref{method}
are given in Table~\ref{tab-ag-like-q}. Lowest-order DF and final SDpT all-order sensitivity coefficients $q$ are given for comparison. The Breit interaction is included and QED is omitted in the calculation of $q$ factors.
Final SDpT all-order transition energies are given for reference. All energy and $q$ values are given relative to the ground state in cm$^{-1}$.
SDpT energies and $q$ coefficients are used to calculate
 enhancement
factors $K=2q/\omega$ given in the last column of the table. For consistency, we use our final
theoretical values of energies to calculate $K$ for all ions considered in this work.  We find that while the correlation correction is very important for accurate calculation of the
transition energies, it only weakly affects the values of $q$. The DF values differ from the final all-order values of $q$ by less than $2\%$ with the
exception of the $4f_{5/2}-4f_{7/2}$ transition in Sm$^{15+}$, where the transition energy is relatively small and correlation contributes 6.3\%.
The enhancement factors are large for the $5s-4f$ transitions for both ions.
 \begin{figure}
  \includegraphics[scale=0.45]{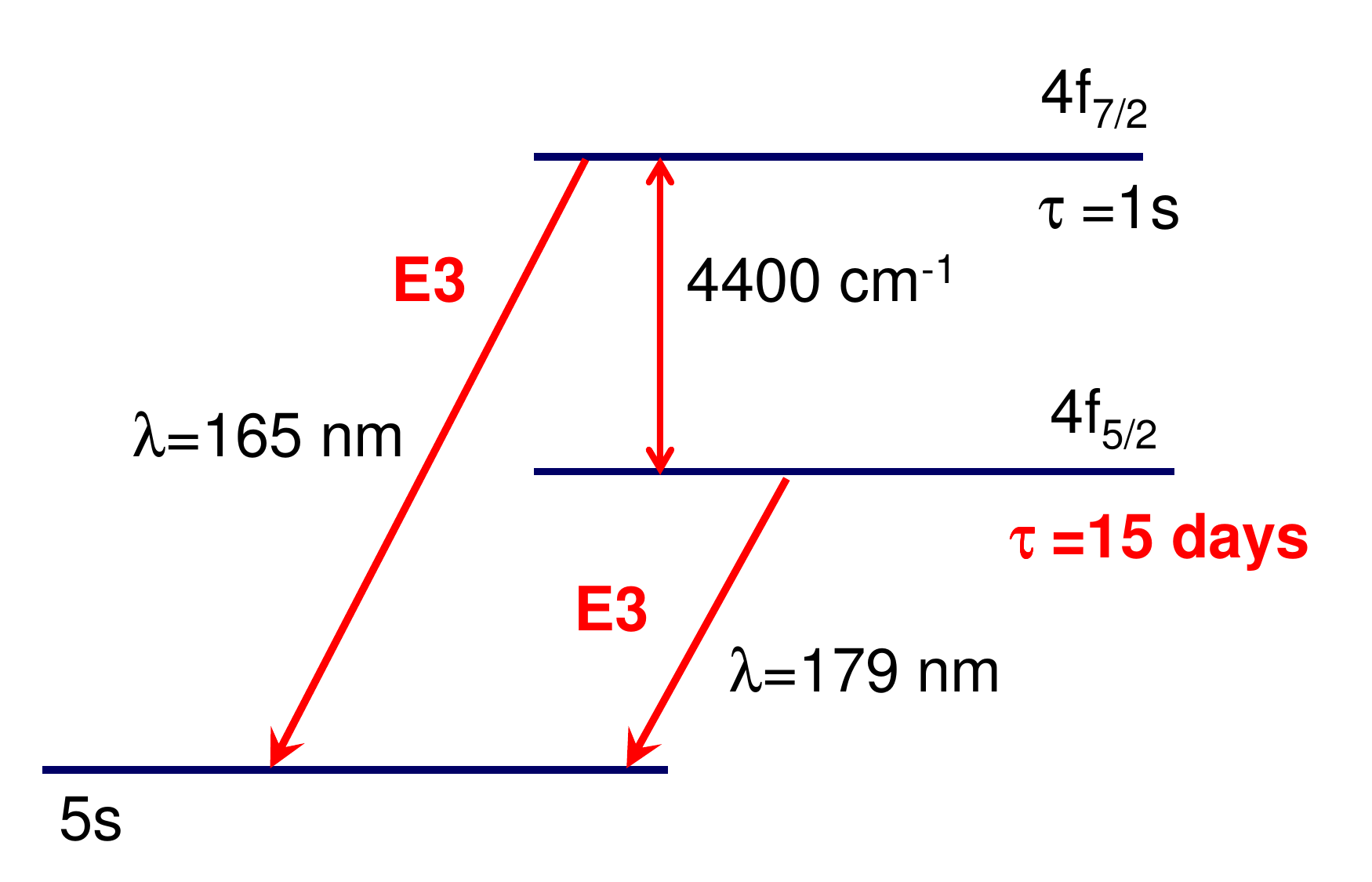}
  \caption{\normalsize{Energy levels and radiative lifetimes of low-lying levels of Ag-like Nd$^{13+}$.}
    \label{fig1}}
\end{figure}

Lowest order ($Z^{\rm DF}$) and all-order ($Z^{\rm SDpT}$)  multipole matrix elements
$Ek$ and $Mk$, $k=1,2,3$ in a.u. and corresponding  transition rates $A$ (in s$^{-1}$) are given
in Table~\ref{tab-mult1}. The transition rates are calculated using the formulas in Section~\ref{method}. Experimental energies from Ref.~\cite{ag-like-81} are used in transition rate calculations. The experimental energies (in cm$^{-1}$) and wavelengths (in nm) from Ref.~\cite{ag-like-81} are listed in Table~\ref{tab-mult1}.
The lifetimes are given in the same row as the level designation for the convenience of presentation.
A single transition gives the dominant contribution to the lifetimes of Nd$^{13+}$ states considered in this work. The strongest transition from the metastable $4f_{5/2}$ level of this ion is E3, resulting in the extremely long lifetime of more than 15 days. Therefore, this system may be considered to have two ground states. The low-lying levels of Nd$^{13+}$ ion and our estimates of  the radiative lifetimes are shown in Fig.~\ref{fig1} for illustration. Long lifetimes of Nd$^{13+}$ ion and large values of $q$ make it particularly attractive candidate for applications considered in this work.

 There are two significant contributions to the lifetime of the $5s$ state in Sm$^{15+}$, $5s-4f_{5/2}$ and $5s-4f_{7/2}$, both of which are E3 transitions. The $5s-4f_{5/2}$ M2 transition gives negligible
contribution. The $5s$ state has also extremely long lifetime, 3.6 days, but has two decay channels.

The sensitivity of transitions in Nd$^{13+}$ and
Sm$^{15+}$ to variation of $\alpha$ as well as uncertainty budget of the atomic clocks
were discussed in detail in Ref.~\cite{DzuDerFla12}. Blackbody radiation shift, Zeeman shift, electric quadrupole shift, and other perturbations affecting clock frequencies were considered in \cite{DzuDerFla12} and the ultimate fractional frequency uncertainty was projected at $10^{-19}$.

\begin{table*}
\caption{\label{tab-in1-like} Energies of In-like ``monovalent'' Cs$^{6+}$, Ba$^{7+}$, Ce$^{9+}$, Pr$^{10+}$, and
Nd$^{11+}$ ions relative to the ground state  in cm$^{-1}$. Results of two different methods, monovalent coupled-cluster SDpT  and CI+all-order,
are given in columns labeled SDpT and CI+all. The CI+all-order results are taken as final.
 Contributions from higher-order Coulomb correlation (difference of the CI+MBPT and CI+all-order calculations),
estimated contributions of  higher partial waves (above $l>6$), and
Breit interaction are given separately in columns HO, Extrap, and Breit, respectively.
Experimental results are from \cite{nist-web} for Cs$^{6+}$ and Ba$^{7+}$ and from \cite{in-like-01} for Ce$^{9+}$.
Differences with experiment are given in cm$^{-1}$ and \% in columns ``Diff.''  and ``Diff.\%''. Estimated uncertainties of
theoretical calculations are given in column ``Unc''.
 Wavelengths for
transitions to the ground state are given in last two columns in nm.}
\begin{ruledtabular}
\begin{tabular}{llrrrrrrrrrrrcc}
\multicolumn{1}{c}{Ion}&
\multicolumn{1}{c}{Level}&
\multicolumn{1}{c}{Expt}&
\multicolumn{1}{c}{SDpT}&
  \multicolumn{1}{c}{Diff.}&  \multicolumn{1}{c}{CI+MBPT}& \multicolumn{1}{c}{HO}&
\multicolumn{1}{c}{Extrap}&
\multicolumn{1}{c}{Breit}&
\multicolumn{1}{c}{CI+all}&
\multicolumn{1}{c}{Unc.}&
\multicolumn{1}{c}{Diff.}&
\multicolumn{1}{c}{Diff.\%}& \multicolumn{1}{c}{$\lambda_\textrm{th}$}&
 \multicolumn{1}{c}{$\lambda_\textrm{expt}$}\\
 \hline    \\[-0.4pc]
 Cs$^{6+}$  &   $5p_{1/2}$&      0&     0  &      0& 0     &       0&      0&      0&      0       &     &    &       &           &  \\
            &   $5p_{3/2}$& 19379 & 19351  &     28& 19733 &    -127&     11&   -245& 19372        &     &  7 &   0.04&516.20     & 516.01   \\
            &   $4f_{5/2}$& 166538& 166851 &   -313& 166341&    1839&   -740&   -995& 166446       &     &  92&   0.06&60.08      &  60.05    \\
            &   $4f_{7/2}$& 167297& 167603 &   -306& 167234&    1787&   -733&  -1103& 167186       &     & 111&   0.07&59.81      &   59.77\\[0.5pc]
  Ba$^{7+}$  &  $5p_{1/2}$&      0&      0 &      0&      0&       0&      0&    0  &      0       &     &    &       &           &        \\
             &  $5p_{3/2}$& 23592 &  23564 &28     & 24020 &-134    &12     &-293   &23605         &     &-13 &-0.05  &423.65     &  423.87\\
             &  $4f_{5/2}$& 137385&  137770&-385   & 137086&2224    &-858   &-1197  &137256        &     &129 &0.09   &72.86      &  72.79\\
             &  $4f_{7/2}$& 138675&  139043&-368   & 138570&2169    &-851   &-1345  &138542        &     &133 &0.10   &72.18      &  72.11\\[0.5pc]
 Ce$^{9+}$  &  $5p_{1/2}$&       0&      0 &      0&      0&     0  &0      &   0   &      0       &     &    &       &          &  \\
            &  $5p_{3/2}$&  33427 &  33406 &     21& 33986 &  -147  &14     &-403   &  33450       & 130 & -23&  -0.07&299.0     & 299.2 \\
            &  $4f_{5/2}$&  54947 &  55419 &   -472& 54601 &  2687  &-1011  &-1595  &  54683       & 220 & 264&   0.48&182.9     & 182.0 \\
            &  $4f_{7/2}$&  57520 &  57968 &   -448& 57441 &  2628  &-1004  &-1830  &  57235       & 310 & 285&   0.50&174.7     & 173.9 \\  [0.5pc]
 Pr$^{10+}$ &  $5p_{1/2}$&        &     0  &       &     0 &     0  &      0&      0&  0           &     &    &       &          &  \\
            &  $4f_{5/2}$&        &  3958  &       & 3471  & 2821   &  -1063&  -1797&3702$^{\rm a}$& 200 &    &       & 2700(140)&\\
            &  $4f_{7/2}$&        &  7276  &       & 7136  & 2761   &  -1057&  -2079&7031$^{\rm a}$& 200 &    &       &  1422(40)&\\
            &  $5p_{3/2}$&        &  39084 &       & 39745 & -154   &   14  &   -464&39141         & 40  &    &       &  255.5(3)&\\  [0.5pc]
 Nd$^{11+}$ &  $4f_{5/2}$&        &       0&       &     0 &     0  &    0  &      0&0             &     &    &       &          &  \\
            &  $4f_{7/2}$&        &  4155  &       & 4566  & -61    &7      &-332   &  4180        &100  &    &       &  2392(60)&\\
            &  $5p_{1/2}$&        &  52823 &       & 53491 & -2916  &1106   &2003   &   53684      &500  &    &       & 186.3(1.7)&\\
            &  $5p_{3/2}$&        &  98175 &       & 99549 & -3076  &1121   &1472   &   99066      &500  &    &       &   100.9(5)&\\
\end{tabular}
\begin{flushleft}
$^{\rm a}$These values are adjusted by 270~cm$^{-1}$
based on the comparison of Ce$^{9+}$ results with experiment.
\end{flushleft}
\end{ruledtabular}
\end{table*}

\begin{table}
\caption{\label{tab-pr-q} Comparison of sensitivity coefficients $q$
in In-like Pr$^{10+}$ ions relative to the ground state evaluated in the lowest-order Dirac-Fock (DF+Breit), SDpT, and CI+all-order
approximation in cm$^{-1}$. The second-column (DF) shows lowest-order results without the Breit interaction. Breit is included in all other calculation.  }
\begin{ruledtabular}
\begin{tabular}{rrrrr}
\multicolumn{1}{c}{Level}&
\multicolumn{1}{c}{DF}&
\multicolumn{1}{c}{DF+Breit}&
\multicolumn{1}{c}{SDpT}&
\multicolumn{1}{c}{CI+all}\\
\hline   \\[-0.4pc]
$5p_{1/2}$   &   0   &   0   &   0   &   0   \\
$4f_{5/2}$   &   75276   &   73494   &   73865   &   73849   \\
$4f_{7/2}$   &   78081   &   76059   &   76803   &   76833   \\
$5p_{3/2}$   &   44552   &   43977   &   44091   &   44098   \\
 \end{tabular}
\end{ruledtabular}
\end{table}

\begin{table}
\caption{\label{tab-in-like-q} Energies and  sensitivity coefficients $q$
for In-like ions relative to the ground state evaluated in the CI+all-order
approximation in cm$^{-1}$; $K=2q/\omega$ is the enhancement
factor. }
\begin{ruledtabular}
\begin{tabular}{rrrrr}
\multicolumn{1}{c}{Ion}&
\multicolumn{1}{c}{Level}&
\multicolumn{1}{c}{Energy}&
\multicolumn{1}{c}{$q$}&
\multicolumn{1}{c}{$K$}\\
\hline   \\[-0.4pc]
Ce$^{9+}$   &  $5p_{1/2}$&  0   &   0   &       \\
    &  $5p_{3/2}$&  33450   &   37544   &   2.2 \\
    &  $4f_{5/2}$&  54683   &   62873   &   2.3 \\
    &  $4f_{7/2}$&  57235   &   65150   &   2.3 \\[0.5pc]
Pr$^{10+}$  &  $5p_{1/2}$&  0   &   0   &       \\
    &  $4f_{5/2}$&  3702    &   73849   &   40  \\
    &  $4f_{7/2}$&  7031    &   76833   &   22  \\
    &  $5p_{3/2}$&  39141   &   44098   &   2.3 \\[0.5pc]
Nd$^{11+}$  &  $4f_{5/2}$&  0   &   0   &       \\
    &  $4f_{7/2}$&  4180    &   3785    &   1.8 \\
    &  $5p_{1/2}$&  53684   &   -85692  &   -3.2    \\
    &  $5p_{3/2}$&  99066   &   -34349  &   -0.7    \\
\end{tabular}
\end{ruledtabular}
\end{table}

\begin{table*}
\caption{\label{tab-mult2} Lowest order ($Z^{\rm DF}$) and all-order ($Z^{\rm SDpT}$)  multipole matrix elements
$E2$ and $M1$,  transition rates $A$ (in s$^{-1}$), and lifetimes in In-like ions. Experimental energies from Refs.~\cite{nist-web,in-like-01} are used for Cs$^{6+}$, Ba$^{7+}$, and
Ce$^{9+}$ ions; theoretical energies from Table~\ref{tab-in1-like} are used for Pr$^{10+}$ and Nd$^{11+}$. Energies (in cm$^{-1}$) and wavelengths (in nm) are listed for reference.   The numbers in brackets represent
powers of 10.}
\begin{ruledtabular}
\begin{tabular}{lllrrrrrr}
\multicolumn{1}{c}{Level}& \multicolumn{2}{c}{Transition}&
\multicolumn{1}{c}{Energy}& \multicolumn{1}{c}{$\lambda$}&
\multicolumn{1}{c}{$Z^{\rm DF}$}& \multicolumn{1}{c}{$Z^{\rm
SDpT}$}& \multicolumn{1}{c}{$A$}&\multicolumn{1}{c}{Lifetime} \\ \hline
 \multicolumn{9}{c}{  In-like Cs$^{6+}$}\\ \\[-0.6pt]
$5p_{3/2}$ &  $5p_{3/2}-5p_{1/2}$&M1 &  19379&  516.0&   1.152&  1.152&  6.510[01]&1.52[-02]~s\\
           &  $5p_{3/2}-5p_{1/2}$&E2 &  19379&  516.0&   3.100&  2.885&  6.371[-01]&\\[0.3pc]
$4f_{5/2}$ &  $4f_{5/2}-5p_{1/2}$&E2 & 166538&  60.05&   2.608&  2.315&  1.282[04]&6.71[-05]~s\\
           &  $4f_{5/2}-5p_{3/2}$&E2 & 147159&  67.95&   1.416&  1.271&  2.080[03]&\\[0.3pc]
$4f_{7/2}$ &  $4f_{7/2}-4f_{5/2}$&M1 &    759&  13175&   1.851&  1.851&  5.053[-03]&1.03[-04]~s\\
           &  $4f_{7/2}-4f_{5/2}$&E2 &    759&  13175&   0.970&  0.784&  2.166[-09]&\\
           &  $4f_{7/2}-5p_{3/2}$&E2 &  147918&  67.61&   3.486&  3.123&  9.668[03]&\\[0.4pc]
 \multicolumn{9}{c}{  In-like Ba$^{7+}$}\\  \\[-0.6pt]
$5p_{3/2}$ &  $5p_{3/2}-5p_{1/2}$& M1 &  23592&  423.9&   1.152&  1.151&  1.174[02]&8.43[-03]~s\\
           &  $5p_{3/2}-5p_{1/2}$& E2 &  23592&  423.9&   2.725&  2.544&  1.324[00]&\\[0.3pc]
$4f_{5/2}$ &  $4f_{5/2}-5p_{1/2}$& E2 & 137385&  72.79&   1.984&  1.770&  2.862[03]&3.13[-04]~s\\
           &  $4f_{5/2}-5p_{3/2}$& E2 & 113793&  87.88&   1.068&  0.963&  3.304[02]&\\[0.3pc]
$4f_{7/2}$ &  $4f_{7/2}-4f_{5/2}$& M1 &   1290&   7752&   1.851&  1.851&  2.480[-02]&6.30[-04]~s\\
           &  $4f_{7/2}-5p_{3/2}$& E2 & 115083&  86.89&   2.635&  2.371&  1.589[ 03]&\\[0.4pc]
 \multicolumn{9}{c}{  In-like Ce$^{9+}$}\\   \\[-0.6pt]
 $5p_{3/2}$ &  $5p_{3/2}-5p_{1/2}$& M1 &  33427&  299.2&   1.151&  1.151&  3.335[02]&3.00[-03]~s\\
           &  $5p_{3/2}-5p_{1/2}$& E2 &  33427&  299.2&   2.175&  2.040&  4.860[00] &\\[0.3pc]
$4f_{5/2}$ &  $4f_{5/2}-5p_{1/2}$& E2 &  54947&  182.0&   1.277&  1.146&  1.229[01] &8.12[-02]~s\\
           &  $4f_{5/2}-5p_{3/2}$& E2 &  21520&  464.7&   0.679&  0.615&  3.260[-02]&\\[0.3pc]
$4f_{7/2}$ &  $4f_{7/2}-4f_{5/2}$& M1 &   2573&   3887&   1.851&  1.851&  1.968[-01]&\textbf{2.18~s}\\
           &  $4f_{7/2}-5p_{3/2}$& E2 &  24093&  415.1&   1.677&  1.518&  2.620[-01]&\\[0.4pc]
  \multicolumn{9}{c}{  In-like Pr$^{10+}$}\\  \\[-0.6pt]
$4f_{5/2}$&$ 4f_{5/2}-5p_{1/2}$&  E2&   3702&     2701&     1.062&     0.955&    1.183[-05] &\textbf{1.0~day}\\[0.3pc]
$4f_{7/2}$&$ 4f_{7/2}-4f_{5/2}$&  M1&   3329&     3004&     1.851&     1.851&    4.260[-01] &\textbf{2.35~s}\\[0.3pc]
$5p_{3/2}$&$ 5p_{3/2}-5p_{1/2}$&  M1&  39141&     255.5&    1.151&     1.151&    5.352[02]  &1.83[-03]~s\\
          &$ 5p_{3/2}-5p_{1/2}$&  E2&  39141&     255.5&    1.969&     1.851&    8.810      &\\
          &$ 5p_{3/2}-4f_{5/2}$&  E2&  35439&     282.2&    0.561&     0.509&    4.058[-01] &\\
          &$ 5p_{3/2}-4f_{7/2}$&  E2&  32110&     311.4&    1.388&     1.258&    1.512      &\\[0.4pc]
\multicolumn{9}{c}{  In-like Nd$^{11+}$}\\ \\[-0.6pt]
 $4f_{7/2}$& $4f_{7/2}-4f_{5/2}$ & M1&  4180&    2392.3&     1.851&     1.851&    8.434[-01]&\textbf{1.19~s} \\[0.3pc]
 $5p_{1/2}$& $5p_{1/2}-4f_{5/2}$ & E2& 53684&     186.3&     0.902&     0.811&     1.641[01]&6.09[-02]~s \\[0.3pc]
 $5p_{3/2}$& $5p_{3/2}-5p_{1/2}$ & M1& 45382&     220.4&     1.150&     1.150&     8.336[02]&8.76[-04]~s \\
           & $5p_{3/2}-5p_{1/2}$ & E2& 45382&     220.4&     1.790&     1.685&     1.530[01]& \\
           & $5p_{3/2}-4f_{5/2}$ & E2& 99066&     100.9&     0.474&     0.430&     4.938[01]& \\
           & $5p_{3/2}-4f_{7/2}$ & E2& 94886&     105.4&     1.173&     1.064&     2.436[02]& \\
  \end{tabular}
\end{ruledtabular}
\end{table*}

\begin{table}
\caption{\label{tab-mult-1-3}
 Comparison of multipole matrix elements in In-like  Pr$^{10+}$.
calculated using SDpT (one-electron) and CI+All (three-electron)
methods.}
\begin{ruledtabular}
\begin{tabular}{rrrrrrr}
\multicolumn{1}{c}{}& \multicolumn{2}{c}{Transition}&
\multicolumn{1}{c}{Energy}& \multicolumn{1}{c}{$\lambda$}&
\multicolumn{1}{c}{$Z^{\rm
SDpT}$}&
\multicolumn{1}{c}{$Z^{\rm CI+all}$}\\
\multicolumn{1}{c}{}& \multicolumn{2}{c}{}& \multicolumn{2}{c}{}&
\multicolumn{1}{c}{One-el.}&
\multicolumn{1}{c}{Three-el.}\\
\hline
 M1&   $4f_{5/2}$&$  4f_{7/2}$&    3329&    3004&        1.85064&    1.85045\\
 E3&   $4f_{5/2}$&$  4f_{7/2}$&    3329&    3004&       0.29479&    0.29157\\
 M1&   $4f_{5/2}$&$  5p_{3/2}$&   35439&     282.2&        0.00012&    0.00006\\
 E2&   $4f_{5/2}$&$  5p_{3/2}$&   35439&     282.2&        0.50920&    0.51700\\
 E2&   $4f_{7/2}$&$  5p_{3/2}$&   32110&     311.4&        1.25759&    1.27820\\
 E2&   $5p_{1/2}$&$  4f_{5/2}$&    3702&     2701&         0.95489&   0.96963\\
 M1&   $5p_{1/2}$&$  5p_{3/2}$&   39141&     255.5&        1.15049&   1.15070  \\
 E2&   $5p_{1/2}$&$  5p_{3/2}$&   39141&     255.5&        1.85071&   1.87010
\end{tabular}
\end{ruledtabular}
\end{table}

\begin{table*}
\caption{\label{tab-in2-like} Energies of In-like ``trivalent'' Sm$^{13+}$ and Eu$^{14+}$ ions
relative to the ground state calculated using the CI+all-order method
(in cm$^{-1}$).
 Contributions from higher-order Coulomb correlation (difference of the CI+MBPT and CI+all-order calculations),
estimated contributions of  higher partial waves (above $l>6$),
Breit interaction, and QED are given separately in columns HO, Extrap, Breit, and QED.
 Estimated uncertainties of
theoretical calculations are given in column ``Unc''.
 Wavelengths for
transitions to the ground state are given in the last  column in nm.}
\begin{ruledtabular}
\begin{tabular}{lrrrrrrrrr}
\multicolumn{1}{c}{Level}&
 \multicolumn{1}{c}{$J$}&
  \multicolumn{1}{c}{CI+MBPT}&
\multicolumn{1}{c}{HO}&
\multicolumn{1}{c}{Extrap}&
\multicolumn{1}{c}{Breit}&
\multicolumn{1}{c}{QED}&
\multicolumn{1}{c}{Final}&
\multicolumn{1}{c}{Unc.}& \multicolumn{1}{c}{$\lambda_\textrm{th}$}\\
 \hline    \\[-0.6pc]
  \multicolumn{9}{c}{ Sm$^{13+}$~~ $5s^24f_{5/2}$ ground state }\\
  \hline      \\[-0.7pc]
$5s^24f      $&$7/2$ &6667 & -62&        7&    -440&      32 &6203 &   100&  1612(28)\\
$4f^25s      $&$7/2$ &21164& 2983&    -1118&   -1626&   -1149 &20254&   940&   494(22)\\
$4f^25s      $&$9/2$ &23606& 2954&    -1117&   -1785&   -1139 &22519&   950&   444(18)\\
$4f^25s      $&$11/2$&27339& 2893&    -1113&   -2097&   -1119 &25904&   980&   386(14)\\
$4f^25s      $&$ 3/2$&30282& 2787&    -1079&   -1599&   -1142 &29249&   900&   342(10)\\
$4f^25s      $&$13/2$&31557& 2827&    -1108&   -2452&   -1097 &29727&  1000&   336(11)\\
$4f^25s      $&$ 5/2$&31906& 2765&    -1079&   -1720&   -1133 &30739&   900&   325(9)\\
$4f^25s     $&$  7/2$&34189& 2730&    -1079&   -1932&   -1122&32786&   920&   305(8)\\
$4f^25s     $&$  9/2$&34418& 2864&    -1102&   -1893&   -1135&33152&   950&   302(8)\\
$4f^25s     $&$  9/2$&36335& 2811&    -1098&   -2048&   -1130&34871&   950&   287(8)\\
$4f^25s     $&$  7/2$&41004& 2755&    -1085&   -1971&   -1130&39572&   930&   253(6)\\
$4f^25s     $&$ 11/2$&42079& 2873&    -1116&   -2387&   -1130&40319&  1000&   248(6)\\
$4f^25s     $&$  5/2$&43341& 2774&    -1082&   -1771&   -1130&42132&   900&   237(5)\\
   \hline    \\[-0.6pc]
    \multicolumn{9}{c}{ Eu$^{14+}~~4f^25s~J=7/2$ ground state}\\
    \hline      \\ [-0.7pc]
$4f^3  $&$ 9/2$&2048  & 3909&    -1074&  -1722&  -1265&   1896 &  1400    &\\
$4f^25s$&$ 9/2$&2785  &  -27&        1&   -166&      9&    2603&  130     &3842(190)\\
$4f^3  $&$11/2$&6610  & 3841&    -1073&  -2103&  -1241&   6034 &  1500    &1657(330)\\
$4f^25s$&$11/2$&7319  &  -91&        5&   -535&     33&    6732&  430     &1485(90)\\
$4f^25s$&$ 3/2$&9824  &  -195&      40&     36&     8&    9713 &  80      &1030(9)\\
$4f^3  $&$13/2$&11316 &   3773&   -1070&  -2506&  -1219&   10294&  160     &971(130)\\
$4f^25s$&$ 5/2$&11720 &   -216&      40&   -245&     16&   11316&  270     &884(20)\\
$4f^25s$&$13/2$&12361 &   -159&      10&   -792&     57&   11477&  400     &871(30)\\
$4f^25s$&$ 9/2$&14060 &    -88&      10&   -297&     20&   13705&  250     &730(13)\\
$4f^25s$&$ 7/2$&14501 &   -253&      40&   -349&     20&   13959&  360     &716(18)\\
$4f^3  $&$15/2$&16072 &   3705&   -1067&  -2918&  -1240&   14553&  1800    &687(75)\\
 \end{tabular}
 \end{ruledtabular}
\end{table*}

\begin{table}
\caption{\label{tab-in2-like-q} Energies and  sensitivity coefficients $q$
for In-like ions relative to the ground state evaluated in the CI+all-order
approximation in cm$^{-1}$; $K=2q/\omega$ is the enhancement
factor. }
\begin{ruledtabular}
\begin{tabular}{lllrrr}
\multicolumn{1}{c}{Ion}&
\multicolumn{1}{c}{Level}&\multicolumn{1}{c}{$J$}&
\multicolumn{1}{c}{Energy}&
\multicolumn{1}{c}{$q$}&
\multicolumn{1}{c}{$K$}\\
\hline   \\[-0.4pc]
Sm$^{13+}$  &   $5s^2 4f      $&$5/2$ &      0       &   0       &       \\
            &   $5s^2 4f      $&$7/2$ &      6203    &   5654    &   1.8 \\
            &   $4f^2 5s      $&$7/2$ &      20254   &   123621  &   12  \\
            &   $4f^2 5s      $&$9/2$ &      22519   &   125397  &   11  \\
            &   $4f^25s       $&$11/2$&      25904   &   128875  &   10  \\
            &   $4f^25s       $&$ 3/2$&      29249   &   124872  &   8.5 \\[0.5pc]
Eu$^{14+}$  &   $4f^25s$&$7/2$       &   0        &   0       &       \\
            &   $4f^3  $&$ 9/2$       &   1896    &   137437  &   145 \\
            &   $4f^25s$&$ 9/2$       &   2603    &   1942    &   1.5 \\
            &   $4f^3  $&$11/2$       &   6034    &   141771  &   47  \\
            &   $4f^25s$&$11/2$       &   6732    &   6293    &   1.9 \\
            &   $4f^25s$&$ 3/2$       &   9713    &   1474    &   0.3 \\
            &   $4f^3  $&$13/2$       &   10294   &   145723  &   28
    \end{tabular}
\end{ruledtabular}
\end{table}

\begin{table*}
\caption{\label{tab-life-sm13}
CI+all-order $Z^{\rm CI+all}$ multipole matrix elements (in a.u.), transition rates $A_r$ (in s$^{-1}$), and
lifetimes $\tau ^{\rm CI+all}$ (in sec) in In-like Sm$^{13+}$ ion.
Energies (in cm$^{-1}$) and wavelengths (in nm) are from Table~\ref{tab-in2-like}.
The numbers in brackets represent powers of 10. CI+all-order matrix elements without RPA correction are listed in column labelled
$Z^{\rm  noRPA}$. }
\begin{ruledtabular}
\begin{tabular}{lllrrrrrrl}
\multicolumn{1}{c}{Level}&
\multicolumn{2}{c}{Transition}&
\multicolumn{1}{c}{}&
\multicolumn{1}{c}{Energy}&
\multicolumn{1}{c}{$\lambda$}&
\multicolumn{1}{c}{$Z^{\rm  noRPA}$}&
\multicolumn{1}{c}{$Z^{\rm  CI+all}$}&
\multicolumn{1}{c}{$A_{r}^{\rm  CI+all}$}&
\multicolumn{1}{c}{$\tau^{\rm  CI+all}$}\\
\hline
$ 5s^24f\ ^2F_{7/2}$ &  $  5s^24f\ ^2F_{5/2}$& $ 5s^24f\ ^2F_{7/2}$&  M1&   6203&  1612&   1.84090&  1.84102&  2.728[+0]& 0.367\\
                     &  $  5s^24f\ ^2F_{5/2}$& $ 5s^24f\ ^2F_{7/2}$&  E2&   6203&  1612&   0.21559&  0.19504&  4.891[-6]&      \\[0.4pc]
$ 5s4f^2\ ^4H_{7/2}$ &  $  5s^24f\ ^2F_{5/2}$& $ 5s4f^2\ ^4H_{7/2}$&  E1&  20254&   493.7&   0.00195&  0.00188&  7.443[+0]& 0.133\\
                     &  $  5s^24f\ ^2F_{7/2}$& $ 5s4f^2\ ^4H_{7/2}$&  E1&  14051&   711.7&   0.00033&  0.00027&  4.949[-2]&      \\[0.4pc]
$ 5s4f^2\ ^4H_{9/2}$ &  $  5s^24f\ ^2F_{7/2}$& $ 5s4f^2\ ^4H_{9/2}$&  E1&  16316&   612.9&   0.00295&  0.00275&  6.636[+0]& 0.141\\
                     &  $  5s4f^2\ ^4H_{7/2}$& $ 5s4f^2\ ^4H_{9/2}$&  M1&   2265&  4415&   3.85348&  3.85331&  4.654[-1]&      \\[0.4pc]
$ 5s4f^2\ ^4H_{11/2}$&  $  5s4f^2\ ^4H_{9/2}$& $ 5s4f^2\ ^4H_{11/2}$& M1&   3385&  2954&   4.46131&  4.46125&  1.735[+0]& 0.576\\[0.4pc]
$ 5s4f^2\ ^4F_{3/2}$ &  $  5s^24f\ ^2F_{5/2}$& $ 5s4f^2\ ^4F_{3/2}$&  E1&  29249&   341.9&   0.00384&  0.00342&  1.484[+2]& 6.74[-3]\\
                     &  $  5s4f^2\ ^4H_{7/2}$& $ 5s4f^2\ ^4F_{3/2}$&  E2&   8995&  1112&   0.45992&  0.41986&  2.906[-4]&      \\[0.4pc]
$ 5s4f^2\ ^4H_{13/2}$&  $  5s4f^2\ ^4H_{11/2}$&$ 5s4f^2\ ^4H_{13/2}$& M1&   3823&  2616&   3.94521&  3.94520&  1.676[+0]& 0.597\\[0.4pc]
$ 5s4f^2\ ^4F_{5/2}$ &  $  5s^24f\ ^2F_{5/2}$& $ 5s4f^2\ ^4F_{5/2}$&  E1&  30739&   325.3&   0.00321&  0.00293&  8.426[+1]&  9.62[-3]\\
                     &  $  5s^24f\ ^2F_{7/2}$& $ 5s4f^2\ ^4F_{5/2}$&  E1&  24536&   407.6&   0.00221&  0.00198&  1.956[+1]&          \\
                     &  $  5s4f^2\ ^4F_{3/2}$& $ 5s4f^2\ ^4F_{5/2}$&  M1&   1490&  6711&   3.02239&  3.02227&  1.358[-1]&     \\[0.4pc]
$ 5s4f^2\ ^4F_{7/2}$ &  $  5s^24f\ ^2F_{5/2}$& $ 5s4f^2\ ^4F_{7/2}$&  E1&  32786&   305.0&   0.00304&  0.00280&  6.993[+1]&  1.20[-2]\\
                     &  $  5s^24f\ ^2F_{7/2}$& $ 5s4f^2\ ^4F_{7/2}$&  E1&  26583&   376.2&   0.00177&  0.00165&  1.292[+1]&          \\
                     &  $  5s4f^2\ ^4H_{7/2}$& $ 5s4f^2\ ^4F_{7/2}$&  M1&  12532&   798.0&   0.17400&  0.17406&  2.011[-1]&          \\
                     &  $  5s4f^2\ ^4F_{5/2}$& $ 5s4f^2\ ^4F_{7/2}$&  M1&   2047&  4885&   3.45771&  3.45759&  3.457[-1]&    \\[0.4pc]
$ 5s4f^2\ ^2H_{9/2}$ &  $  5s^24f\ ^2F_{7/2}$& $ 5s4f^2\ ^2H_{9/2}$&  E1&  26949&   371.1&   0.01033&  0.00948&  3.565[+2]&  2.78[-3]\\
                     &  $  5s4f^2\ ^4H_{7/2}$& $ 5s4f^2\ ^2H_{9/2}$&  M1&  12898&   775.3&   0.68327&  0.68350&  2.704[+0]&    \\[0.4pc]
$ 5s4f^2\ ^4F_{9/2}$ &  $  5s^24f\ ^2F_{7/2}$& $ 5s4f^2\ ^4F_{9/2}$&  E1&  28668&   348.8&   0.00512&  0.00461&  1.014[+2]&  9.61[-3]\\
                     &  $  5s4f^2\ ^4H_{7/2}$& $ 5s4f^2\ ^4F_{9/2}$&  M1&  14617&   684.1&   0.47303&  0.47324&  1.887[+0]&          \\
                     &  $  5s4f^2\ ^4H_{9/2}$& $ 5s4f^2\ ^4F_{9/2}$&  M1&  12352&   809.6&   0.31732&  0.31751&  5.125[-1]&          \\
                     &  $  5s4f^2\ ^4F_{7/2}$& $ 5s4f^2\ ^4F_{9/2}$&  M1&   2085&  4796&   2.17866&  2.17854&  1.160[-1]&     \\[0.4pc]
$ 5s4f^2\ ^2G_{7/2}$ &  $  5s^24f\ ^2F_{5/2}$& $ 5s4f^2\ ^2G_{7/2}$&  E1&  39572&   252.7&   0.00997&  0.00971&  1.481[+3]&  5.31[-4]\\
                     &  $  5s^24f\ ^2F_{7/2}$& $ 5s4f^2\ ^2G_{7/2}$&  E1&  33369&   299.7&   0.00721&  0.00649&  3.966[+2]&        \\
                     &  $  5s4f^2\ ^4H_{7/2}$& $ 5s4f^2\ ^2G_{7/2}$&  M1&  19318&   517.7&   0.36671&  0.36685&  3.271[+0]&         \\
                     &  $  5s4f^2\ ^4H_{7/2}$& $ 5s4f^2\ ^2G_{7/2}$&  E2&  19318&   517.7&   0.05726&  0.05278&  1.049[-4]&         \\
                     &  $  5s4f^2\ ^4H_{9/2}$& $ 5s4f^2\ ^2G_{7/2}$&  M1&  17053&   586.4&   0.31361&  0.31357&  1.644[+0]&     \\[0.4pc]
$ 5s4f^2\ ^2H_{11/2}$&  $  5s4f^2\ ^4H_{9/2}$& $ 5s4f^2\ ^2H_{11/2}$& M1&  17800&   561.8&   0.20453&  0.20464&  5.309[-1]&   0.207    \\
                     &  $  5s4f^2\ ^4H_{13/2}$&$ 5s4f^2\ ^2H_{11/2}$& M1&  10592&   944.1&   0.61358&  0.61320&  1.004[+0]&            \\
                     &  $  5s4f^2\ ^2H_{9/2}$& $ 5s4f^2\ ^2H_{11/2}$& M1&   7167&  1395&   1.62570&  1.62592&  2.188[+0]&            \\
                     &  $  5s4f^2\ ^4F_{9/2}$& $ 5s4f^2\ ^2H_{11/2}$& M1&   5448&  1836&   1.66373&  1.66384&  1.006[+0]&       \\[0.4pc]
$ 5s4f^2\ ^2F_{5/2}$ &  $  5s^24f\ ^2F_{5/2}$& $ 5s4f^2\ ^2F_{5/2}$&  E1&  42132&   237.3&   0.01158&  0.01101&  3.062[+3]&  3.19[-4]  \\
                     &  $  5s^24f\ ^2F_{7/2}$& $ 5s4f^2\ ^2F_{5/2}$&  E1&  35929&   278.3&   0.00242&  0.00214&  7.140[+1]&            \\
                     &  $  5s4f^2\ ^4F_{3/2}$& $ 5s4f^2\ ^2F_{5/2}$&  M1&  12883&   776.2&   0.43435&  0.43453&  1.815[+0]&            \\
\end{tabular}
\end{ruledtabular}
\end{table*}

 \section{In-like ions}
 \label{in}
 In-like ions have two more valence electrons in comparison with Ag-like ions, and in general are considered to be trivalent systems.
 However, the states with $5s^2 nl$ valence configurations above $1s^2 2s^2 2p^6 3s^2 3p^6 3d^{10} 4s^2 4p^6 4d^{10}$ core
 may be considered to be  monovalent with $5s^2$ shell included into the core. These ions may be treated with both monovalent coupled-cluster
 SDpT all-order method and many-electron CI+all-order method. We use both of these approaches and
 compare their accuracy. The accuracy of these methods for neutral Tl and In has been recently discussed in Refs.~\cite{SafMaj13,SafSafPor13}. The trivalent $4f^3$ and $4f^25s$ configurations in Sm$^{13+}$
 and Eu$^{14+}$ can only be treated with the CI+all-order method.

 There are two level crossings of interest for the present work in In-like isoelectronic sequence, $5p-4f$ and $4f-5s$.
 The first one happens for Pr$^{10+}$ and Nd$^{11+}$ and leads to change of level order from   $5p$, $4f$ to $4f$, $5p$.
 Pr$^{10+}$ represents a particularly attractive case where both $4f_j$ levels are located between the $5p_{1/2}$ and $5p_{3/2}$
fine structure multiplet, making $4f_{5/2}$ a very long-lived metastable level.

 Energies of In-like ``monovalent'' Cs$^{6+}$, Ba$^{7+}$, Ce$^{9+}$, Pr$^{10+}$, and
Nd$^{11+}$ ions relative to the ground state  are given in Table~\ref{tab-in1-like} in cm$^{-1}$. We calculate the energies for
all three  Cs$^{6+}$, Ba$^{7+}$, Ce$^{9+}$ ions where the experimental values are available to understand the trends of the
difference with experiment so we can improve the  values and reduce the uncertainty of the  Pr$^{10+}$ energies.
Results of two different methods, monovalent coupled-cluster SDpT  and CI+all-order,
are given in columns labeled SDpT and CI+all. Difference with experiment is given in cm$^{-1}$ and \% in columns ``Diff.''
and ``Diff.\%''. The table clearly illustrates that the CI+all-order method gives the results in better
agreement with experiment.
Therefore, the CI+all-order results are taken as final for all five ions presented in Table~\ref{tab-in1-like}.

 In order to evaluate the accuracy of our values, we carried out several calculations which allowed us to separate the effect of higher orders
of MBPT, Breit interaction, and contributions of higher partial waves. QED correction is small for transitions with no $5s$ state, and is omitted.
The contributions of the
higher-orders is evaluated as the difference of the CI+all-order and CI+MBPT results. The Breit contribution is
calculated as the difference of the results with and without the inclusion of this effect. The contribution of the higher $(l>6)$
partial waves (labeled ``Extrap'') is estimated to be equal to the contribution of the $l=6$ partial wave following our empiric rule obtained for Ag-like ions. The contribution of the $l=6$ partial wave is obtained as the difference of two calculations where all intermediate sums in the all-order and MBPT terms are restricted to $l_{\rm max}=6$  and $l_{\rm max}=5$.
The resulting contributions from higher-order Coulomb correlation (difference of the CI+MBPT and CI+all-order calculations),
estimated contributions of  higher partial waves (above $l>6$), and
Breit interaction are given separately in columns HO, Extrap, and Breit of Table~\ref{tab-in1-like}.
 The final theoretical results are listed in ``CI+all'' column.

The experimental results are from \cite{nist-web} for Cs$^{6+}$ and Ba$^{7+}$ and \cite{in-like-01} for Ce$^{9+}$. Wavelengths for
transitions to the ground state are given in last two columns in nm.
Estimated uncertainties of
theoretical calculations are given in column ``Unc''. We use Ba$^{7+}$ reference ion to estimate the uncertainties of the
Ce$^{9+}$ calculations using the approaches described in the previous sections. The uncertainty is estimated as the sum of the following: (1) difference of the theoretical and experimental energies for the reference ion (Ba$^{7+}$) and (2) difference in the sum of all four corrections between the reference and the current ion, Ce$^{9+}$.
The resulting uncertainties for the $4f$ states are very close to the actual differences with experiment, while the uncertainty of the $5p$ state is significantly overestimated.

The energies of the $4f$ levels of Pr$^{10+}$ are very difficult to calculate accurately since these are very close to the ground $5p_{1/2}$
 state. The one-electron removal energies of the $5p_{1/2}$ and $4f_{5/2}$ states are $-1.3\times10^6$~cm$^{-1}$, and these values cancel to 99.7~\% when two energies are subtracted to obtain the ~\textit{ab initio} transition energy of 3432~cm$^{-1}$. Meanwhile, all
 of the corrections (HO, Breit, and Extrap) are large, 1000-3000~cm$^{-1}$ and partially cancel each other. Studying the trends of the difference between theory and experiment for three previous ions of the sequence shows that the discrepancy  somewhat increases for heavier ions, which probably results from rapid increase of actual removal energies. Therefore, we adjust our \textit{ab initio} values 3432~cm$^{-1}$ and
6761~cm$^{-1}$
for the $4f_{5/2}$ and $4f_{7/2}$ states, respectively, by the difference of the Ce$^{9+}$ results
 with experiment, i.e. 270~cm$^{-1}$. The change in sum of all three corrections between Ce$^{9+}$ and Pr$^{10+}$ is only 120-170~cm$^{-1}$. Therefore, we estimate the uncertainty in these energies to be on the order of $200$~cm$^{-1}$.
 The uncertainty in the $5p_{3/2}$ energy is taken to be 40~cm$^{-1}$ based on the accuracy of this energy for Ce$^{9+}$ ion.

The sensitivity coefficients $q$ for Pr$^{10+}$  obtained in the lowest-order
Dirac-Fock (DF+Breit), SDpT, and CI+all-order approximations are given in
Table~\ref{tab-pr-q} in cm$^{-1}$. The second-column (DF) shows lowest-order
results without the Breit interaction. Breit is included in all other
calculations. Just as in the case of Ag-like ions, the correlation effect on the
values of $q$ is small, 0.3-1.0\%. The Breit interaction contributes from
-1.3\% to -2.7\%. The differences between the coefficients $q$ calculated
using the CI+all-order and SDpT methods are negligible. The final CI+all-order
sensitivity coefficients $q$ for ``monovalent'' In-like Ce$^{9+}$, Pr$^{10+}$,
and  Nd$^{11+}$ ions are given in Table~\ref{tab-in-like-q} together with the
corresponding CI+all-order transition energies and $K$
 enhancement
factors.

\begin{figure}
  \includegraphics[scale=0.45]{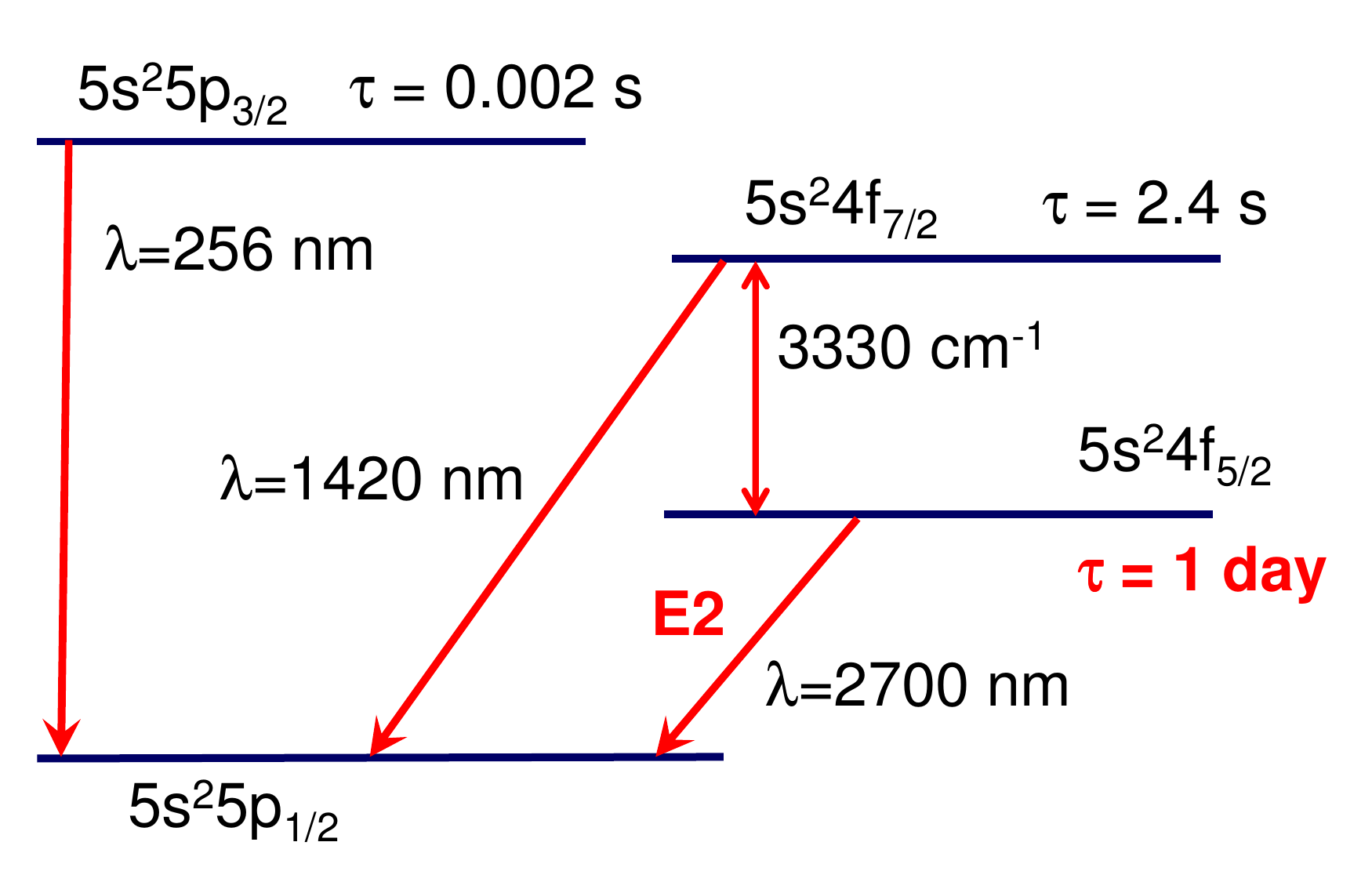}
  \caption{\normalsize{Energy levels and radiative lifetimes of low-lying levels of In-like Pr$^{10+}$.}
    \label{fig2}}
\end{figure}

Lowest-order ($Z^{\rm DF}$) and all-order ($Z^{\rm SDpT}$)  multipole matrix elements
$E2$ and $M1$,  transition rates $A$ (in s$^{-1}$), and lifetimes in In-like ions are listed in Table~\ref{tab-mult2}. The lifetimes are given in the same row as the level designation for the convenience of presentation. Experimental energies from Refs.~\cite{nist-web,in-like-01} are used for Cs$^{6+}$, Ba$^{7+}$, and
Ce$^{9+}$ ions; theoretical energies from Table~\ref{tab-in1-like} are used for Pr$^{10+}$ and Nd$^{11+}$. Energies (in cm$^{-1}$) and wavelengths (in nm) are listed for reference.  Multipole matrix elements are
evaluated by the SDpT method. We have verified that all other transitions give negligible contributions to the lifetimes.
Comparison of multipole matrix elements in In-like  Pr$^{10+}$ calculated using the  
 SDpT (one-electron) and CI+All (three-electron)
methods is given in Table~\ref{tab-mult-1-3}. The values calculated by both methods are in excellent agreement.
The low-lying levels of Pr$^{10+}$ ion and our estimates of  the radiative lifetimes are shown in Fig.~\ref{fig2} for illustration.

 The order of levels changes again for Sm$^{13+}$, where $5s4f^2$
 configuration becomes the closest to the ground $5s^24f_{j}$ fine-structure
 multiplet. This leads to a very interesting level structure with a metastable
 $5s4f^2$ $J=7/2$ level in the optical transition range to both ground and
 excited  $5s^24f_{7/2}$ levels of the fine-structure multiplet. The second
 level crossing, $5s-4f$, leads to further change of the level order for
 Eu$^{14+}$, where $5s4f^2$ becomes the ground state and the $4f^3$ becomes the
 first excited level. We note that these levels are very close and the
 uncertainty of our calculations is comparable to the energy interval.
 Therefore, it might be possible that $4f^3$ $J=9/2$ is a ground state
 configuration. The previous In-like reference ions, such as Ce$^{9+}$ cannot
 be used to establish the accuracy of the calculations for Sm$^{13+}$ and
 Eu$^{14+}$ due to completely different set of low-lying configurations. The
 25\% of all four corrections added in quadrature is used to estimate
 uncertainty for all levels of Sm$^{13+}$ and $4f^3$ levels of Eu$^{14+}$. In
 the case of the $5s^24f$ fine structure multiplet energy levels of
 Eu$^{14+}$, we take the average of 25\% estimate and sum of the all four
 corrections as an uncertainty. We note that this is the first time that the
 CI+all-order method was applied to such complicated configurations as $4f^3$
 and no benchmark comparisons with experiment exist for such states.
 Therefore, experimental measurement of Eu$^{14+}$ will serve as an excellent
 benchmark of the method accuracy  that will allow to further develop the
 methodology for more complicated systems with partially filled $nf$ shells.

The final CI+all-order sensitivity coefficients $q$ for ``trivalent'' In-like
Sm$^{13+}$ and Eu$^{14+}$ ions are given in Table~\ref{tab-in2-like-q}
together with the corresponding CI+all-order transition energies and $K$
enhancement factors.

The
CI+all-order $Z^{\rm CI+all}$ multipole matrix elements (E1, E2,
E3, M1, M2, and M3), transition rates $A_r$ (in s$^{-1}$), and
lifetimes $\tau ^{\rm CI+all}$ (in sec) in In-like Sm$^{13+}$ ion are listed in Table~\ref{tab-life-sm13}.
Energies from Table~\ref{tab-in2-like} used for evaluation of the matrix elements and transition rates are
given for reference. Multipole matrix
elements are evaluated in the CI+all-order approximations (a.u.).
The numbers in brackets represent powers of 10.
The CI+all-order matrix elements calculated without random-phase-approximation (RPA)
 correction are listed in column labelled
$Z^{\rm  noRPA}$. In such a calculation, ``bare'' $Ek$ and $Mk$ operators are used instead of the
effective transition operators (for example electric-dipole
$D^{\textrm{eff}})$. While RPA correction is significant for E1 and E2 transitions, it is small for M1
transitions between the levels of the fine-structure multiplet.

We find that the lifetimes of the Sm$^{13+}$ levels are relatively small, less then 1~sec, making it less attractive for our applications of interest. Nevertheless, shorter lifetimes will make locating the transitions easier so
this ion may be used as a benchmark for further improvement of the theory. If measurements are carried out in
this ion, it may be possible to use the resulting comparison to further improve the theory for other ions.

The case of 
Eu$^{14+}$ is similar and does not appear to have all features for the clock development. 
While uncertainty of our data is large, present results place  the first level too close to the
 ground state to be practically useful.
The next level is of the same configuration as the ground state and
 has small value of $q$. Finally, $4f^3~ J=11/2$ level is 
 short lived since it can decay via the $M1$ transition  to $4f^3~ J=9/2$ level below. 
However, measurement of this ion energy levels would be very useful as the benchmark for other systems.

\section{Conclusion}

We carried out detailed high-precision study of Ag-like Nd$^{13+}$ and Sm$^{15+}$
and In-like Ce$^{9+}$, Pr$^{10+}$, Nd$^{11+}$, Sm$^{13+}$, and Eu$^{14+}$
highly-charged ions for future  experimental studies aimed at the development
of ultra-precise atomic clocks and search for $\alpha$-variation.  The
energies of Nd$^{13+}$, Sm$^{15+}$ and In-like Ce$^{9+}$ ions were found to be
in excellent agreement with experiment. The energies, transition wavelengths,
electric- and magnetic-multipole reduced matrix elements, lifetimes, and
sensitivity coefficients to $\alpha$-variation $q$ and $K$ were calculated.
Several methods were developed to evaluate uncertainties of the results.
Particulary interesting cases for experimental exploration were highlighted.

\section*{Acknowledgement}
We thank C. W. Clark, C. Monroe, J. Tan, Yu. Ralchenko, and P. Beiersdorfer for useful discussions. M.S.S. thanks School of Physics at the University of New South Wales, Sydney, Australia for hospitality and acknowledges support from Gordon Godfrey Fellowship, UNSW. This work was supported in part by US NSF Grant No.\ PHY-1212442. M.G.K. acknowledges support from RFBR Grant No.\ 14-02-00241.
The work was partly supported by the Australian Research Council.


\end{document}